\documentclass[aps,10pt,nofootinbib,amsmath,prd,twocolumn,showpacs,superscriptaddress,groupedaddress]{revtex4-1}

\usepackage[american]{babel}
\usepackage{amsfonts}
\usepackage{amsmath}
\usepackage{amssymb}
\usepackage{slashed}
\usepackage{xcolor}
\usepackage{bm}
\usepackage{float}
\usepackage[utf8]{inputenc}
\usepackage[macrosonly]{chet}
\usepackage{nicefrac}
\usepackage{hyperref}
\usepackage{url}

\usepackage{lmodern}
\usepackage{tikz}
\usetikzlibrary{decorations.pathmorphing}

\tikzset{graviton/.style={decorate, decoration={snake, segment length=2mm, amplitude=0.6mm}}}

\newcommand{\dd}{{\rm d}}
\newcommand{\tr}{{\rm tr}}

\begin{document}

\title{Physical running in conformal gravity and higher derivative scalars}

\author{Diego Buccio}
\email{dbuccio@sissa.it}
\affiliation{International School for Advanced Studies, Via Bonomea 265,
34134 Trieste, Italy and INFN - Sezione di Trieste, Trieste, Italy}

\author{Luca Parente}
\email{luca.parente@phd.unipi.it}
\affiliation{Universit\`a di Pisa and INFN - Sezione di Pisa, Largo Bruno Pontecorvo 3, 56127 Pisa, Italy}

\author{Omar Zanusso}
\email{omar.zanusso@unipi.it}
\affiliation{Universit\`a di Pisa and INFN - Sezione di Pisa, Largo Bruno Pontecorvo 3, 56127 Pisa, Italy}

\begin{abstract}
%
We compute the physical running of a general higher derivative scalar coupled to a nondynamical metric and of higher derivative Weyl invariant gravity with a dynamical metric in four dimensions.
In both cases, we find that the physical running differs from the $\mu$-running of dimensional regularization because of infrared divergences
which are present in amplitudes also at large momenta, differently from what happens in standard two derivative theories.
We use the higher derivative scalar as a toy-model to elaborate on the properties of the conformal limit in relation to the trace anomaly.
The physical running of higher derivative Weyl gravity, while different from the $\mu$-running, remains asymptotically free, suggesting that the model is a viable completion of Einstein's gravity, at least from the point of view of its renormalization group properties.
\end{abstract}

\maketitle

\section{Introduction}\label{sect:intro}

The renormalization group (RG) is one of the most powerful tools in modern quantum field theory. Despite its wide diffusion and the important results that it allowed to achieve, it is sometimes not completely clear what the relations among its different implementations are.
In the particle physics community, the RG was introduced by Gell-Mann and Low \cite{Gell-Mann:1954yli} as a smart way to reabsorb via a redefinition of the coupling constants the large logarithms of kinematical variables that emerge in loop corrections to scattering amplitudes and that can potentially disrupt the perturbative expansion in the high-energy regime. In the following, we will refer to the dependence of the theory's parameters on energy scales of a process as ``physical running'' in accordance with this procedure.

In another interpretation of the RG given by Wilson \cite{Wilson:1973jj}, which was later extended to high energy physics thanks to dimensional regularization, the renormalization group describes how the parameters of an effective theory depend on a UV cutoff and the way in which it regularizes divergences.
In order to regularize UV divergences, an unphysical energy scale must be introduced in the theory, independently of the preferred renormalization scheme. Physics must be independent of this scale, but renormalized amplitudes explicitly depend on it. Hence, renormalized couplings must run with the scale to obtain physically meaningful results.
We call this approach $\mu$-running, from the usual name given to the unphysical scale $\mu$ introduced by dimensional regularization. In two derivative theories, the $\mu$- and physical runnings are equivalent in the high energy limit, however the same is not true in general.

Higher dimensional operators that are at least quadratic in the curvatures are generated by loop corrections in gravitational theories, making general relativity nonrenormalizable \cite{tHooft:1974toh, Goroff:1985th}.
On the other hand, higher derivative quantum field theories are less UV divergent with respect to standard two derivative ones, so quadratic gravity is often suggested as a possible solution to this problem \cite{Stelle:1976gc}.
However, the improved UV behavior is accompanied by new IR divergences. With a quartic propagator, many loop integrals become IR divergent, hence an infrared mass must be introduced as a regulator. In two derivative theories these IR divergences can be treated via the definition of IR safe observables because they emerge only in the presence of soft or collinear particles, but in higher derivative theories IR divergences are present also for large momenta, thus such a cancellation is impossible. The direct consequence is the appearance in the amplitudes of large logarithms of kinematical variables that are independent of the UV regulator. It has been recently discussed the possibility that, due to these IR divergences, the beta functions given by the physical definition of the running may differ from the $\mu$-running in some higher derivative theories \cite{Buccio:2023lzo, Donoghue:2023yjt}.
With this idea in mind, the dependence on logarithms of the D'Alembertian operator in the one-loop effective action of quadratic gravity was calculated in Ref.~\cite{Buccio:2024hys}. The result was a new set of beta functions with the surprising feature of allowing asymptotic freedom without tachyonic modes in the spectrum. In this work, we extend this approach to the cases of a higher derivative scalar field coupled to gravity and to conformal gravity in four spacetime dimensions. In both cases, we find a discrepancy between the physical running and the $\mu$-running, which was calculated in Refs.~\cite{Barvinsky:1985an} and \cite{Antoniadis:1992xu,deBerredo-Peixoto:2003jda}.

In detail, in section \ref{sect:hd-scalar} we discuss the different notions of running couplings in the quantum effective action, then we compute the contributions to the one-loop effective action of the higher derivative scalar field, with particular emphasis on the conformal limit of the theory.
In section \ref{sect:hd-conf-grav} we adapt the calculation of \cite{Buccio:2024hys} to the case of conformal gravity and find the new beta function produced by the physical running scheme.
In section \ref{sect:conclusions} we summarize the results and comment on future perspectives.

\section{Higher-derivative scalar field}\label{sect:hd-scalar}

In this section, we regard the metric as a nondynamical field, which could be seen simply as the source of the energy-momentum tensor.
The most general quadratic action of a dimensionless scalar field coupled to the metric in four dimensions is
\begin{eqnarray}\label{eq:action-hds}
        S_{\rm hds}[\varphi] &=& \frac{1}{2} \int \dd^4 x \sqrt{g} \varphi {\cal D}_4 \varphi 
        \,, \\
        {\cal D}_4 \varphi &=& \Box^2 \varphi
        + \nabla_\mu \Bigl( \bigl( \xi_1 R^{\mu\nu} + \xi_2 g^{\mu\nu} R\bigr) \nabla_\nu \varphi\Bigr)
        + E \varphi
        \,, \nonumber \\
        E &=& \lambda_1 C^2 + \lambda_2 R_{\mu\nu}R^{\mu\nu} + \lambda_3 R^2 + \lambda_4 \Box R\,, \nonumber
\end{eqnarray}
where $C^2 = C_{\mu\nu\rho\theta}C^{\mu\nu\rho\theta}$ is the square of Weyl's tensor. The differential operator ${\cal D}_4$ is written in such a way that it is manifestly self-adjoint. In the limit $\lambda_i=0$ the action $S_{\rm hds}[\varphi]$ is shift-invariant, i.e., invariant under $\varphi\to \varphi +c$ for constant $c$, which can be seen easily integrating by parts the first two terms. Another relevant limit is $\lambda_i=0$, $\xi_1=2$ and $\xi_2=-\frac{2}{3}$, for which ${\cal D}_4$ becomes the Weyl covariant Paneitz-Riegert operator and the action itself becomes conformal invariant \cite{Fradkin:1982xc, Riegert:1984kt, Paneitz}. In the conformal limit, the trace of the classical variational energy-momentum tensor of $S_{\rm hds}[\varphi]$ is zero going on-shell using the equations of motion of $\varphi$ (i.e., ${\cal D}_4\varphi=0$), in agreement with the Noether identities of Weyl symmetry. {In the Sec.~\ref{sect:conformal} we return to the conformal limit and refer to the scalar action in this particular choice of the parameters as $S_{\rm c}[\varphi]$.}

The effective action in curved space is obtained by integrating $\varphi$ in the path-integral. Since $\varphi$ appears quadratically in $S_{\rm hds}[\varphi]$, the effective action depends only on the metric, and at one loop it is
\begin{equation}\label{eq:effective-action}
    \begin{split}
        \Gamma &= S_{\rm hds} + \frac{1}{2} \tr \log S_{\rm hds}^{(2)}
        \,.
    \end{split}
\end{equation}
In asymptotically flat spaces\footnote{{In the asymptotic region the metric can be decomposed in $g_{\mu\nu}=\delta_{\mu\nu}+h_{\mu\nu}$, with $h_{\mu\nu}$ small at infinity as boundary condition. Consequently, we expect boundary terms of $\Gamma$ to be negligible in this configuration.}} the regularized effective action can be parametrized in such a way that it resums derivatives, {as shown in detail in Refs.~\cite{Barvinsky:1990up,Codello:2012kq} and consistent with our computations below,} giving
\begin{eqnarray}\label{eq:action-form-factors}
\begin{split}
        \Gamma &= S + \int \dd^4 x \sqrt{g} \Bigl\{
        C_{\mu\nu\alpha\beta} f_{\lambda}(\Box;\mu^2,m^2) C^{\mu\nu\alpha\beta}
        \\&
        \qquad \qquad  + R f_{\xi}(\Box;\mu^2,m^2) R
        \Bigr\} +{\cal O}({\cal R}^3)
        \,,
\end{split}
\end{eqnarray}
where the form-factors $f_i$ for $i=\lambda,\xi$ depend on $\mu$, which is the scale introduced to regulate ultraviolet divergences (i.e., from dimensional regularization) and on $m$, which is an infrared mass that regulates infrared divergences. In momentum space and using dimensional regularization, $d=4-\epsilon$, the UV divergent part of the form factors is
\begin{equation}\label{eq:form-factors-div-part}
    \begin{split}
        f_{i,{\rm div}}(p^2;\mu^2,m^2)
        &= 2 b_i \frac{\mu^\epsilon}{\epsilon}
        \,,
    \end{split}
\end{equation}
{ which is guaranteed by the fact that ultraviolet divergences must be local.}
The scales are separated as $m^2\ll p^2\ll \mu^2$ and the $b_i$s are some constants that depend on the original action \eqref{eq:action-hds}.
The UV divergence can be subtracted with standard renormalization and results in $\mu$-dependent couplings and corresponding beta functions by taking the logarithmic $\mu$-derivative. By construction, the $\mu$-running depends on the coefficients $b_i$.

After the subtraction of the ultraviolet divergences, the renormalized form-factors have the general structure for $\epsilon\to 0$ 
\begin{equation}\label{eq:form-factors-finite-part}
    \begin{split}
        f_{i,{\rm ren}}(p^2;\mu^2,m^2)
        &= 
        b_i \log(\mu^2/ m^2) + c_i \log(p^2/m^2)
        \,, 
    \end{split}
\end{equation}
which add to renormalized combinations of the couplings at the scale $\mu$, and this form can always be achieved by combining the logarithms for dimensionless $f_i$s. { This form is expected from power counting in loop diagrams. All terms in the action \eqref{eq:action-hds} are marginal, hence they can only produce contributions to other marginal operators. Since $C^2$ and $R^2$ are marginal too,  when scales are largely separated their form factors cannot contain powers of $p$, but only logarithms. This is also confirmed by the explicit computations below.}

Notice that the coefficients of the two logarithms of Eq.~\eqref{eq:form-factors-finite-part} are generally different, $c_i\neq b_i$. This implies that the integration of the $\mu$-running does not resum the large logarithms in $p^2$, unless $b_i=c_i$.\footnote{%
It is known also through explicit computations that two-derivative scalars and vectors, as well as spinor fields, give effective actions in curved space for which $b_i=c_i$. See for example Refs.~\cite{Franchino-Vinas:2018gzr} and references therein.
}
In order to have a running that resums the large logarithms, the first term can be absorbed through a finite subtraction. The result is a coupling renormalized at the scale $m$, with a form factor that depends only on $p^2$ through $\log(p^2/m^2)$. At this point, by moving the renormalization point from $m$ to $\bar p$ close to $p$, we carry out the desired task. Identifying $p=\left|p\right|$ with the scale of some physical process,
the logarithmic derivative with respect to $p$ gives the physical running, which, by construction, depends only on $c_i$.

\subsection{Computational approaches}\label{sect:methods}

The relevant information for the renormalization of $\Gamma$ can be extracted from the two-point functions with external ``graviton'' legs (recall that the metric is nondynamical in this section, so the leg corresponds to a functional derivative with respect to the background).
At $1$-loop we have two contributions
\begin{equation} \label{eq:diagrams}
\begin{split}
       \Gamma^{(2)}_{1-{\rm loop}} & = 
        -\frac{1}{2} \Bigl\{ 2~
        \begin{tikzpicture}[baseline=-.1cm]
        \draw[draw=black, graviton] (.5,0) -- (1.2,0);
        \draw[draw=black, graviton] (-1.2,0) -- (-.5,0);
        \draw (0,0) circle (.5cm);
        \filldraw [gray!50] (.5,0) circle (2.5pt);
        \draw (.5,0) circle (2.5pt);
        \filldraw [gray!50] (-.5,0) circle (2.5pt);
        \draw (-.5,0) circle (2.5pt);
        \end{tikzpicture} ~ 
        -~
        \begin{tikzpicture}[baseline=-.2cm]
        \draw[draw=black, graviton] (0,-0.5) -- (0.7,-0.7);
        \draw[draw=black, graviton] (-0.7,-0.7) -- (0,-0.5);
        \draw (0,0) circle (.5cm);
        \filldraw [gray!50] (0,-.5) circle (2.5pt);
        \draw (0,-.5) circle (2.5pt);
        \end{tikzpicture}
        ~ \Bigr\}
     \end{split}
\end{equation}
{ which is obtained by applying two functional derivatives to Eq.~\eqref{eq:effective-action}},
with incoming momentum $p_\mu$ entering from the left, referred to as ``bubble'' and ``tadpole'' from now on. Denoting $q_\mu$ the integrated momentum, we have that both diagrams depend on $p_\mu$ through the vertices, but only the bubble depends nontrivially on $p_\mu$ through the two scalar propagators.

For the example of this section, the internal lines represent scalar propagators, which are higher derivative and need to be regulated in the infrared in $d=4$. Naively, the bubble contains the product of the propagators $q^{-4}(q+p)^{-4}$, which we regulate in two distinct ways. We choose either
\begin{equation}\label{eq:regularization-ir-v1}
    \begin{split}
        \frac{1}{q^4(q+p)^4} \longrightarrow \frac{1}{(q^2+m^2)^2((q+p)^2+m^2)^2}\,,
    \end{split}\nonumber
\end{equation}
or, as done in Ref.~\cite{Buccio:2024hys},
\begin{equation}\label{eq:regularization-ir-v2}
    \begin{split}
        \frac{1}{q^4(q+p)^4} \longrightarrow \frac{1}{q^2(q^2+m^2)(q+p)^2((q+p)^2+m^2)}\,.
    \end{split}\nonumber
\end{equation}
The tadpole must then be regulated in the same way as the bubble, but in the limit $p_\mu=0$. In all computations we have checked that both infrared regularizations lead to the same result, suggesting that the $\log(m^2)$ contributions to the form factors are universal.
In the ultraviolet, the diagrams can be either regulated with dimensional regularization, as described above, or with an ultraviolet cutoff. We have chosen the former for simplicity, given that we are renormalizing dimensionless couplings.

The explicit computation of the form factors can be performed in two equivalent ways. The first approach is the one discussed in Ref.~\cite{Buccio:2024hys}, in which the background metric is expanded as $g_{\mu\nu}=\delta_{\mu\nu}+f_{\mu\nu}$
and the covariant expression of $\Gamma$ is reconstructed from the contraction of \eqref{eq:diagrams} with two copies of $f_{\mu\nu}$ in momentum space. The approach is described in detail in Ref.~\cite{Buccio:2024hys}, so we do not repeat the details here.

A second approach follows a different strategy, used in Ref.~\cite{Codello:2012kq} for a similar computation, and involves the comparison of \eqref{eq:diagrams} with the projection of the second variation of \eqref{eq:action-form-factors} using the decomposition in spin-projectors in momentum space. Denoting the transverse
spin-$1$ projectors as $P_{\mu\nu} = \delta_{\mu\nu} - p^{-2} ~ p_\mu p_\nu$,
we define the transverse-traceless ($TT$) and scalar spin-$2$ projectors in $d=4$ as
\begin{equation}\label{eq:projectors-spin-2}
    \begin{split}
        H^{\alpha\beta}_{\mu\nu}
        = P^{\alpha}_{(\mu} P^{\beta}_{\nu)}
        -\frac{1}{3} P_{\mu\nu} P^{\alpha\beta}
        \,,\quad 
        S^{\alpha\beta}_{\mu\nu}
        = \frac{1}{3} P_{\mu\nu} P^{\alpha\beta}\,.
    \end{split}
\end{equation}
The complete decomposition includes two more projectors that we do not need for this presentation \cite{Stelle:1976gc}. The second variation of \eqref{eq:effective-action} with respect to the metric in the flat space limit and in momentum space is expressed using the projectors as
\begin{equation}\label{eq:second-variation-action-ansatz}
    \begin{split}
        \left.\frac{\delta^{2} \Gamma}{\delta g_{\mu\nu} \delta g_{\alpha\beta}}\right|_{\rm flat}
        &= 2 f_\lambda H_{\mu\nu\alpha\beta} + 12 f_\xi S_{\mu\nu\alpha\beta} + \cdots
        \,,
    \end{split}
\end{equation}
where the dots hide the other spin-projectors.
It follows that Eq.~\eqref{eq:second-variation-action-ansatz} can be compared with the explicit computation of Eq.~\eqref{eq:diagrams}
to extract the form factors $f_\lambda$ and $f_\xi$ as functions of $p^2$.
The comparison allows for the separate determination of the coefficients of the dimensional poles $b_i$ as well as the $c_i$ that multiply the $\log(p^2)$ terms. All the results provided in the rest of the paper have been tested in multiple ways, using different UV and IR regularizations
to check for their universality.

\subsection{Beta functions}\label{sect:betas-scalar}

To extract meaningful runnings for the actual couplings, we write the renormalized effective action schematically as local plus nonlocal parts
\begin{equation}\label{eq:renormalized-action-local}
    \begin{split}
        \Gamma_{\rm ren} = & \int \dd^4 x \sqrt{g} \Bigl\{
        \frac{1}{2\lambda }C^2+\frac{1}{\xi} R^2 
        \Bigr\} + \Gamma_{\rm nl}[g]
        \,,
    \end{split}
\end{equation}
where $\Gamma_{\rm nl}[g]$ includes the renormalized nonlocal contributions given in  Eq.~\eqref{eq:form-factors-finite-part}. The renormalized couplings of Eq.~\eqref{eq:renormalized-action-local}
can be arranged to resum either the $\log(\mu^2)$ or the $\log(p^2)$ logarithms, but not both. As a consequence, we have the two different notions of RG running. The $\mu$-running of the couplings is
\begin{equation}\label{eq:beta-functions-mu-running}
    \begin{split}
        \beta^{\rm s}_{\frac{1}{\lambda}} = -2 \mu \frac{\dd f_\lambda}{\dd \mu}  = -4 b_\lambda
        \,, \quad
        \beta^{\rm s}_{\frac{1}{\xi}} &= - \mu \frac{\dd f_\xi}{\dd \mu}  = -2 b_\xi
        \,.
    \end{split}
\end{equation}
The physical running is obtained equivalently and, in this notation, just amounts to replacing the coefficients $b_i$ with $c_i$
\begin{equation}\label{eq:beta-functions-physical-running}
    \begin{split}
        \beta^{\rm s,ph}_{\frac{1}{\lambda}} = -4 p^2 \frac{\dd f_\lambda}{\dd p^2}  = -4 c_\lambda
        \,, \quad
        \beta^{\rm s,ph}_{\frac{1}{\xi}} &= - 2 p^2 \frac{\dd f_\xi}{\dd p^2}  = -2 c_\xi
        \,.
    \end{split}
\end{equation}
The difference between the two runnings can only be due to the operators $\lambda_i$. This has two explanations: on the one hand, the tadpole diagram is logarithmically divergent only if there are no derivatives of the fluctuating field in the interaction vertex; on the other hand, infrared divergences in the bubble integral occur only if in the numerator there are no powers of the momentum transported by both propagators. These two conditions for the appearance of $\log(m^2)$ terms in the one-loop corrected effective action are equivalent to require $E\ne0$ in the action (\ref{eq:action-hds}). The operators related to $\lambda_1$, $\lambda_2$ and $\lambda_3$ are quadratic in curvatures, so they can contribute at quadratic order in the metric's fluctuation only to the $\mu$-running via a tadpole diagram. In contrast, $\Box R$ is linear in curvature, so it can contribute via bubbles to the physical running, but, if inserted in a tadpole, it will give a total derivative. Then, $E=0$ means that the fluctuating field has no effective mass and in this case no discrepancy in the two definitions of running coupling is expected.

The explicit computation is consistent with the above expectations. It reveals for the $\mu$-running
\begin{equation}\label{eq:beta-functions-mu-running-actual}
    \begin{split}
        (4\pi)^2\beta^{\rm s}_{\frac{1}{\lambda}} &= \frac{1}{30} +\frac{\xi_1}{24}(\xi_1-4) -2\lambda_1 -\lambda_2
        \,,
        \\
        (4\pi)^2\beta^{\rm s}_{\frac{1}{\xi}} &= \frac{1}{18} 
        +\frac{\xi_1}{18}+\frac{5\xi^2_1}{72}+\frac{\xi_2}{3}+\frac{\xi_1 \xi_2}{2}+\xi^2_2
        \\&
        -\frac{2\lambda_2}{3}-2\lambda_3
        \,,
    \end{split}
\end{equation}
which agrees with Ref.~\cite{Barvinsky:1985an},
and for the physical running
\begin{eqnarray}\label{eq:beta-functions-physical-running-actual}
        (4\pi)^2\beta^{\rm s,ph}_{\frac{1}{\lambda}} &=& \frac{1}{30} +\frac{\xi_1}{24}(\xi_1-4) 
        \,,
        \\
        (4\pi)^2\beta^{\rm s,ph}_{\frac{1}{\xi}} &=& \frac{1}{18} 
        +\frac{\xi_1}{18}+\frac{5\xi^2_1}{72}+\frac{\xi_2}{3}+\frac{\xi_1 \xi_2}{2}+\xi^2_2
        -2\lambda_4^2
        \,.\nonumber
\end{eqnarray}
A consistency check shows that the difference between the two runnings is always equal to the logarithmic derivative with respect to $m^2$, in agreement with the expression given in Eq.~\eqref{eq:form-factors-finite-part}.

\subsection{Conformal limit and the trace-anomaly}\label{sect:conformal}

The variational energy-momentum tensor of $S_{\rm c}$ is defined as $T^{\mu\nu} = -\frac{2}{\sqrt{g}} \frac{\delta S_{\rm c}}{\delta g_{\mu\nu}}$. Using the equation of motion ${\cal D}_4 \varphi=0$ of the scalar, it is easy to show that diffeomorphisms invariance implies that it is conserved, $\nabla_\mu T^{\mu\nu}=0$.
In the conformal limit, $\lambda_i=0$, $\xi_1=2$ and $\xi_2=-\frac{2}{3}$, the variational energy-momentum tensor of $S_{\rm c}$ is also traceless, $T=T^\mu{}_\mu=0$.
Quantum mechanically we have that the path-integral induces an anomaly which in $d=4$ has the general form
\begin{equation}\label{eq:trace-anomaly}
    \begin{split}
        \langle T \rangle = \frac{1}{(4\pi)^2}\Bigl\{ b \, C^2 + a \, E_4
        \Bigr\}
        \,,
    \end{split}
\end{equation}
as dictated by the Wess-Zumino integrability condition \cite{Osborn:1991gm,Deser:1993yx}, where $E_4=R_{\mu\nu\alpha\beta}^2-4R_{\mu\nu}^2 +R^2$
is the Euler density scalar. This implies, for example, that there is no independent $R^2$ term, besides the one in $E_4$. In the above formula, we have discarded a ``trivial'' $\Box R$ anomaly, which can be eliminated by including $R^2$ in $S_{\rm hds}$ when the metric is \emph{not} dynamical and there are no self-interactions of the field $\varphi$.

Using the Callan-Symanzik equation of $\Gamma$ and the fact that $\langle T^{\mu\nu}\rangle = -\frac{2}{\sqrt{g}}\frac{\delta \Gamma}{\delta g_{\mu\nu}}$ for either RG scale, a general argument relates the coefficients of the anomaly with the beta functions of the couplings \cite{Osborn:1991gm}. We expect that, in the conformal limit,
\begin{equation}\label{eq:trace-anomaly-beta}
    \begin{split}
        \langle T \rangle = \frac{1}{2} \beta_{\frac{1}{\lambda}} C^2 + \beta_{\frac{1}{\xi}}R^2
        + \frac{a}{(4\pi)^2} E_4
        \,,
    \end{split}
\end{equation}
where the coefficient $a$ is not determined by our computation given that we are working with two-point functions in asymptotically flat spacetime.

The expressions \eqref{eq:trace-anomaly} and \eqref{eq:trace-anomaly-beta} for $\langle T\rangle$ should be compatible in the overlapping regimes of validity.
Taking into account the fact that we compute the RG on asymptotically flat spacetimes, that is, $E_4=0$, we have that compatibility requires that $\beta_{\frac{1}{\xi}}=0$
in the conformal limit. Fortunately, this is verified by both the $\mu$-running and the physical beta functions given above in Eqs.~\eqref{eq:beta-functions-mu-running-actual} and \eqref{eq:beta-functions-physical-running-actual}, respectively.

Furthermore, the trace anomaly is an observable, in the sense that we can construct identities among renormalized $n$-point functions that are constrained by the form of $\langle T \rangle$. Since the anomaly is an observable {we have that $a$ and $b$ are scheme-independent constants characterizing the underlying conformal field theory in the flat-space limit \cite{Komargodski:2011vj}}. Thus it may be tempting to ask whether the observable coefficients should be determined by the $\mu$-running or by the physical beta functions, given that the Callan-Symanzik equation can be formulated with either.

The answer is actually simpler: in the conformal limit, the $\mu$-running and the physical running do coincide, in agreement with the expectation that the conformal limit should be completely scaleless. {In other words, the scale $m$ decouples in the conformal limit and only $\log(p^2/\mu^2)$ survives in Eq.~\eqref{eq:form-factors-finite-part}}. In fact,
\begin{equation}\label{eq:beta-functions-conformal-limit}
    \begin{split}
        (4\pi)^2\beta^{\rm s,conf}_{\frac{1}{\lambda}} = -\frac{2}{15}
        \,, \qquad
        (4\pi)^2\beta^{\rm s,conf}_{\frac{1}{\xi}} = 0
        \,.
    \end{split}
\end{equation}
Combining everything together we have that the conformal higher derivative scalar has the anomaly
\begin{equation}\label{eq:trace-anomaly-scalar-final}
    \begin{split}
        \langle T \rangle = \frac{1}{(4\pi)^2} \Bigl\{-\frac{1}{15} C^2 + a E_4 \Bigr\}
        \,,
    \end{split}
\end{equation}
where the so-called $b$-anomaly coefficient of $C^2$ agrees with the literature \cite{Asorey:2003uf, Boyle:2021jaz}, while the $a$-anomaly coefficient is not determined by our beta function because we are limited to two-point functions in asymptotically flat spacetime.\footnote{
In fact, in the flat space limit and for nondynamical metric Eq.~\eqref{eq:diagrams} is precisely renormalizing $ \int \dd^4 x \langle T_{\mu\nu}(x) T_{\alpha\beta}(0) \rangle {\rm e}^{i p\cdot x}$, where $T_{\mu\nu}$ is seen as the composite operator sourced by $g_{\mu\nu}$. The coefficients of the anomaly can be related to it in both broken and unbroken phases of conformal symmetry \cite{Komargodski:2011vj}.
}
In order to compute $a$ from physical correlators, it would be necessary to use either $3$- or $4$-point functions, depending on the approach \cite{Osborn:1991gm,Komargodski:2011vj}. However, we know already that $a=\frac{7}{90}$ from standard covariant methods using the heat-kernel expansion \cite{Asorey:2003uf}.

One final observation is that the $\mu$- and physical runnings given in Eqs.~\eqref{eq:beta-functions-mu-running-actual} and \eqref{eq:beta-functions-physical-running-actual} do coincide in the more general limit $\lambda_i=0$
in which the higher derivative action \eqref{eq:action-hds} is shift-invariant \cite{Safari:2021ocb}. Notice that the action requires an integration by parts
in order to be manifestly shift-invariant (i.e., to depend only on $\partial\varphi$). These models admit shift-invariant interactions
and, while naively nonunitary, they have received renewed attention in attempts to generalize the notion of unitarity \cite{Tseytlin:2022flu, Holdom:2023usn}. Furthermore, shift-symmetry plays a role in the construction of a natural virial current which is a signature of a theory that is \emph{scale}-but-not-conformal invariant \cite{Gimenez-Grau:2023lpz}, { which could be a speculative explanation on why the two runnings coincides. Another possible explanation comes from Ref.~\cite{Delzescaux:2023rgm}, where a shift-invariant higher derivative $O(N)$ scalar model is rewritten in terms of a vector field whose kinetic term has $2$-derivative and propagates only the longitudinal mode. If this reformulation is possible in all shift-invariant theories, no difference between runnings can emerge, since one-loop beta functions are universal in $2$-derivative theories.}

\section{Higher-derivative conformal gravity}\label{sect:hd-conf-grav}

In this section, we consider the metric as a dynamical field and study the physical beta functions of conformal gravity. In this theory, the graviton is the only field taken into account and the action reads
\begin{equation}\label{eq:action-cg}
    \begin{split}
        S_{\rm cg}[g_{\mu\nu}]= & \int \dd^4 x \sqrt{g} \Bigl\{
        \frac{1}{2\lambda }C^2-\frac{1}{\rho} E_4 
        \Bigr\}
        \,,
    \end{split}
\end{equation}
which is the most general gravitational action invariant under Weyl transformations. Unlike quadratic gravity \cite{Stelle:1976gc}, only the transverse-traceless ($TT$) part of the metric fluctuations propagates here. 
To define a quantum effective action, we split the metric in a background part $\bar{g}_{\mu\nu}$ and a quantum fluctuation $h_{\mu\nu}$ via a linear relation
\begin{equation}\label{eq:linear_split}
    \begin{split}
       g_{\mu\nu}=\bar{g}_{\mu\nu}+h_{\mu\nu}
        \,.
    \end{split}
\end{equation}
The gauge arbitrariness due to Weyl invariance can be explicitly fixed by projecting the quantum fluctuation on its traceless part, then we have to treat the gauge freedom coming from diffeomorphisms invariance. To handle it, we choose the background gauge 
\begin{equation}\label{eq:gauge}
    \begin{split}
      F_\mu=\bar\nabla^\lambda h_{\lambda\mu}+\beta\bar\nabla_\mu h
        \,,
    \end{split}
\end{equation}
where $h=h^\mu{}_\mu$ and $\bar\nabla$ is the covariant derivative with respect to $\bar{g}$,
and enforce it by adding to the action the gauge fixing term and the action of the Faddeev-Popov ghost $c_\mu$
\begin{eqnarray}\label{eq:GF+FP}
       S_{\rm GF+FP}&=&-\frac{1}{2\alpha}\int \dd^4 x \sqrt{\bar g} \Bigl\{ F_\mu Y^{\mu\nu} F_\nu
       \\
       &&+ i \bar c_\mu Y^{\mu\nu} [ \bar g_{\nu\rho} \bar\Box +(2\beta+1)\bar\nabla_\nu \bar\nabla_\rho +\bar R_{\nu\rho} ] c^\rho
       \Bigr\}\,.\nonumber
\end{eqnarray}
In higher derivative theories it is more convenient to use a higher derivative gauge-fixing condition, so we take
\begin{equation}\label{eq:gaugeY}
    \begin{split}
      Y_{\mu\nu}=\bar g_{\mu\nu}\bar\Box+\gamma\bar\nabla_\mu\bar\nabla_\nu-\delta\bar\nabla_\nu\bar\nabla_\mu
        \,.
    \end{split}
\end{equation}
The leading fourth order part of the kinetic term of the traceless graviton can be reduced to the minimal form $\bar\Box^2$ by a smart choice of gauge fixing parameters \cite{Ohta:2013uca}
\begin{equation}\label{eq:GFparameters}
    \begin{split}
      \alpha=\lambda\,, \qquad \beta=-\frac14 \,, \qquad \gamma=-2 \,, \qquad \delta=1
        \,.
    \end{split}
\end{equation}

The effective action of quadratic gravity is obtained by integrating out $h_{\mu\nu}$ in the path-integral. Also in this case the effective action is a function of the background metric with structure (\ref{eq:renormalized-action-local}) and at one loop it is given by
\begin{equation}\label{eq:effective-action-CG}
    \begin{split}
        \Gamma = S_{\rm cg}[\bar{g}] + \frac{1}{2} \tr \log S_{\rm cg+GF}^{(2)}|_{\bar{g}}-\tr \log \Delta_{\rm gh}-\frac12\tr \log Y
        \,,
    \end{split}
\end{equation}
where
\begin{equation}\label{eq:FP}
    \begin{split}
      \Delta_{\rm gh}= \delta^\mu_\nu\bar\Box+(2\beta+1)\bar\nabla^\mu\bar\nabla_\nu+\bar R^\mu{}_\nu
        \,.
    \end{split}
\end{equation}
The $\mu$-running of conformal gravity was originally calculated in Ref.~\cite{Fradkin:1981iu} and later on reproduced in Refs.~\cite{Antoniadis:1992xu,deBerredo-Peixoto:2003jda}, and it results in
\begin{equation}\label{eq:mu-beta-functions-conformal-gravity}
    \begin{split}
        (4\pi)^2\beta^{\rm cg}_{\frac{1}{\lambda}} = \frac{199}{15}
        \,, \qquad
        (4\pi)^2\beta^{\rm cg}_{\frac{1}{\rho}} = -\frac{137}{60}
        \,.
    \end{split}
\end{equation}

The operators $\Delta_{\rm gh}$ and $Y$ are second order, hence their effect on the physical running coincides with the $\mu$-running. On the other hand, $S_{\rm cg+GF}^{(2)}|_{\bar{g}}$ is a higher derivative operator, so there can be a difference between the two runnings. The computation of the dependence of form factors on external momenta can be performed following the procedure introduced in Ref.~\cite{Buccio:2024hys} and outlined briefly in Sec.~\ref{sect:methods}. We calculate the one-loop corrections to the two-point functions of the background metric perturbation $f_{\mu\nu}$ and match them with the second order perturbation of $C^2$ around flat spacetime to reconstruct the covariant structure of the one-loop effective action.
The physical beta function for $\frac 1\lambda$ turns out to be
\begin{equation}\label{eq:cg-beta-functions-physical-running}
    \begin{split}
        (4\pi)^2\beta^{\rm cg, ph}_{\frac{1}{\lambda}} = \frac{93}{5}
        \,.
    \end{split}
\end{equation}
One can immediately see that in this case, despite the conformal symmetry, the two runnings are different, but their qualitative behavior is preserved. In particular, we have asymptotic freedom for $\lambda>0$, which depends only on the sign of the beta function.
{ We expect the physical running to be independent of the gauge parameters \eqref{eq:GFparameters} as does the $\mu$-running \cite{Avramidi:1985ki}, but this ought to be proven either in general or by a future direct computation as we also remark in the conclusions.}

\subsection{Further properties of the physical running}\label{sect:hd-conf-grav-further}

It was observed for the first time in Ref.~\cite{Fradkin:1981iu} that the difference in the $\mu$-running of $\frac1\lambda$ between quadratic gravity and conformal gravity is equal to the contribution of two free scalar field or, equivalently, of one higher derivative free scalar.  This discrepancy is due to the conformal mode of the graviton, which is not dynamical in conformal gravity \cite{Salvio:2017qkx}. The difference may be seen also as coming from a partial gauge fixing of Weyl invariance \cite[Sect.~V]{Martini:2024tie}.
Comparing the result \eqref{eq:cg-beta-functions-physical-running} with the physical running of quadratic gravity in Ref.~\cite{Buccio:2024hys}, one immediately sees that the difference is no longer equal to the contribution of two free scalar fields. In fact, $\beta^{\rm ph}_{\frac{1}{\lambda}}$ receives more contributions from infrared large logarithms. For example, the mixed diagram containing one traceless and one scalar fluctuation generates corrections to the term independent of $\xi$ in the beta function, schematically 
\begin{equation} \label{eq:diagrams-h-hTT-bubble}
\begin{split}
        \left.\Delta \beta^{\rm ph}_{\frac{1}{\lambda}}\right|_{\xi=0} ~ \log(p^2/m^2)
        ~ \subset ~
        \begin{tikzpicture}[baseline=-.1cm]
        \draw[draw=black, graviton] (.5,0) -- (1.2,0);
        \draw[draw=black, graviton] (-1.2,0) -- (-.5,0);
        \draw[draw=black, graviton] (0,0) circle (.5cm);
        \filldraw [gray!50] (.5,0) circle (2.5pt);
        \draw (.5,0) circle (2.5pt);
        \filldraw [gray!50] (-.5,0) circle (2.5pt);
        \draw (-.5,0) circle (2.5pt);
        \draw[] (0,.6) node[above] { ${\scriptstyle h^{TT}_{\mu\nu}}$};
        \draw[] (0,-.5) node[below] { ${\scriptstyle h^{\mu}{}_\mu}$};
        \end{tikzpicture} 
         \,,
     \end{split}
\end{equation}
which explains why the difference is actually expected.

Moving on to another point, recall that the trace-anomaly as considered in Sec.~\ref{sect:conformal} is well-defined only for a nondynamical metric which acts as a source to the energy-momentum tensor, in which case the anomaly has to satisfy appropriate integrability conditions \cite{Osborn:1991gm}. In the case of a dynamical metric for the conformal invariant theory, there is no natural notion of energy-momentum tensor, unless one considers a pseudotensor. Using $SU(N)$ Yang-Mills gauge theories as guidance, in practice we need to ensure that there is no anomaly at RG fixed points, or else the gauge-invariance of the theory is broken by quantum effects other than the RG. In the case of conformal gravity, we have that $\lambda \to 0^+$ is an asymptotically free fixed point for either runnings \eqref{eq:mu-beta-functions-conformal-gravity} and \eqref{eq:cg-beta-functions-physical-running}. We thus expect that the theory is conformal in the UV, similarly to gauge theories, and argue that there is no inconsistency between \eqref{eq:mu-beta-functions-conformal-gravity} and \eqref{eq:cg-beta-functions-physical-running}, as long as both runnings result in asymptotic freedom. Notice that, in the case of gauge theories, to formally prove conformal invariance it is necessary to work with local couplings, or with the parametrization of the action such that the asymptotically free coupling appears as interaction, rather than as global normalization of $F_{\mu\nu}^2$ \cite{Collins:1976yq}. We may expect the same for conformal higher derivative gravity, with the additional complication of having to deal with an energy-momentum pseudotensor.

\section{Conclusions}\label{sect:conclusions}

We have discussed two four-derivative models that corroborate the idea that the physical RG running, i.e., the running with respect to an energy scale of an amplitude, differs from the running of dimensional regularization because of infrared divergences.

The first model that we have considered is a general quadratic scalar coupled to a nondynamical metric, which could be regarded as a toy model that is simple to dissect. In this model, it is possible to discuss with relative simplicity the role of each interaction in shaping the difference between the two runnings. We have also confirmed the fact that in the conformal limit, in which the action becomes Weyl invariant, the infrared divergences decouple \cite{Elizalde:1994sn,Elizalde:1994nz}, making the two runnings coincide and produce the standard charges when used to determine the trace anomaly. One interesting feature that we have noticed is that the two runnings coincide for a more general, shift-invariant, subset of theories. If this is a manifestation of a general fact, it would be interesting to explore the general implication that shift-invariance has in ensuring that the two notions of RG are the same. A possible route would be to explore the ``trace-anomaly'' of a nonconformal theory in the sense discussed in Refs.~\cite{Duff:1993wm,Casarin:2018odz, Ferrero:2023unz,Bertolini:2024vwu}.

The second model that we have considered is the one of Weyl invariant gravity in which the metric is dynamical and includes only a transverse-traceless propagating mode. 
In this case, we have observed that, while different, both runnings have negative beta functions and lead to asymptotic freedom, suggesting that Weyl gravity is a viable UV complete theory, { at least at one loop}, although the problem of unitarity remains open (several solutions have been proposed, e.g., Refs.~\cite{Anselmi:2018ibi,Mannheim:2020ryw,Donoghue:2021eto, Salvio:2015gsi}).
Importantly, we have argued that the two runnings can in fact be different, even if the gravitational theory is Weyl invariant, as long as they both give rise to asymptotic freedom.
{
Actually, we could argue that asymptotic freedom is truly determined only by the physical running, which is the one linked to observable properties of amplitudes because it resums $\log(p^2)$, so our computation gives
a new important piece of evidence in favor of Weyl gravity being UV complete.} To substantiate our findings it would be important to verify whether the physical beta functions are gauge-independent as expected, which is a more complex task that we have not yet endeavored.


\end{document}